# External Periodic Driving of Large Systems of Globally Coupled Phase Oscillators

T. M. Antonsen Jr., R. T. Faghih, M. Girvan, E. Ott and J. Platig

Institute for Research in Electronics and Applied Physics
University of Maryland, College Park, MD 20742

Large systems of coupled oscillators subjected to a periodic external drive occur in many situations in physics and biology. Here the simple, paradigmatic case of equal-strength, all-to-all sine-coupling of phase oscillators subject to a sinusoidal external drive is considered. The stationary states and their stability are determined. Using the stability information and numerical experiments, parameter space phase diagrams showing when different types of system behavior apply are constructed, and the bifurcations marking transitions between different types of behavior are delineated. The analysis is supported by results of direct numerical simulation of an ensemble of oscillators.

**There are several physical and biological systems that consist of a large number of oscillators coupled together. In some cases, as for example in the case of the circadian rhythms, the system of oscillators is acted upon by a periodic drive (the daily variation in sunlight). In this paper, we consider a simple model of such a situation. With sufficient drive, the system of oscillators becomes entrained to the drive frequency and produces regular rhythms. If the drive is too weak, or if the frequency is too different from the average frequency of the free-running oscillators, the external forcing may fail to entrain the system and lead to irregular rhythms.**

**I. Introduction**

There are many situations where the behavior of a large system of coupled oscillators is of interest [1]. Some representative examples are the following. (1) The heart contains many oscillating pacemaker cells that synchronize to regulate cardiac beating [2]. (2) Both plants [3] and animals [4] regulate their daily circadian rhythms through coupling of oscillatory units (e.g., for the case of mammals, neurons contained in the superchiasmatic nucleus within the brain's hypothalamus [4]). (3) Systems of coupled, oscillatory, chemically reacting agents have been observed to synchronize [5]. (4) Josephson junctions arranged in series in an electrical circuit behave as coupled oscillators [6].

A paradigmatic model for such situations was proposed by Kuramoto, who considered the case of all to all, equal strength coupling of simple phase oscillators [6] (oscillators whose state is completely specified by their phase angle)**.** In this model a group of initially unsynchronized oscillators can spontaneously synchronize if the coupling among the oscillators is sufficiently strong.

An important modification of this situation is the case in which the system of coupled oscillators is subjected to an external periodic drive. For example, in the case of circadian rhythm, the system of coupled biological oscillators is acted upon by the daily variations of sunlight experienced by the organism. Several natural questions for such a situation arise. Does the external drive entrain the oscillator system? How does the answer to this question depend on parameters (e.g., the detuning between the drive frequency and the average oscillator frequency, the strength of the driving signal and the strength of the coupling among individual oscillators)? How does the entrainment reestablish itself after some disturbance to the system (for example, as in jet lag)? Here, by entrained we mean that an average over the oscillator states behaves periodically with the frequency of the drive signal; i.e. the oscillator system is coherently frequency locked to the drive.

In order to address questions of entrainment it is useful to have illustrative models. In this spirit, an obvious simple model is that of the Kuramoto model modified by the presence of a sinusoidal driving,

$$\frac{d\theta_i(t)}{dt} = \omega_i + \frac{K}{N}\sum_j \sin(\theta_j - \theta_i) + M_0 \sin(\Omega t + \alpha - \theta_i).  \tag{1}$$

Here $\theta_i$ is the phase of oscillator $i$, $\omega_i$ is the natural frequency of oscillator $i$, $N$ is the number of oscillators with $N \gg 1$ ($i = 1, 2, ..., N$), $K$ is the coupling strength between oscillators, and $M_0$, $\Omega$, and $\alpha$ are the amplitude, frequency, and phase of the external drive ($M_0 = 0$ for the original Kuramoto model [7]). The frequencies $\omega_i$ are randomly chosen from a probability distribution function (PDF) $G'(\omega)$, and we are interested in the $N \to \infty$ limit. This problem has been previously considered by Sakaguchi [8], who has obtained an equation for possible entrained equilibria of Eq. (1), and he has also examined solutions of (1) numerically. In addition to the problem posed in Eq. (1) first treated by Sakaguchi [8], a generalization to include network coupling has been made in Ref. 9 (i.e., $K$ in (1) is replaced by a coupling matrix $K_{ij}$). Here, however, we will be concerned with the original Sakaguchi problem. Our focus will be on finding entrained equilibria for the system (1), and determining their stability. The stability of the Kuramoto equilibrium ($M_0 = 0$) has recently been shown by Morillo and Strogatz [10]. The driven equilibria ($M_0 \neq 0$) can be either stable or unstable depending on the frequency and strength of the driving term. For the driven case we obtain a closed form solution of the stability problem, which takes a particularly simple form for a distribution of oscillator frequencies in the form of a Lorentzian (Sec. II). The solution of the linear stability problem gives a basic understanding of the behavior of the system. In particular, whether an entrained equilibrium of the system represents an attractor depends on its stability. Further, examination of the stability of the equilibria helps to characterize the bifurcations that mark the transitions between different dynamical behaviors as system parameters are varied (Sec. III).

## II. ANALYSIS OF PHASE LOCKED STATES

Our analysis proceeds under the assumption that the number $N$ of individual oscillators is large and that the ensemble of oscillators can be described by a continuous distribution function $f(\theta', \omega, t)$, where $(2\pi)^{-1} f d\theta' d\omega$ is the fraction of oscillators at time $t$ in the area $d\theta' d\omega$ centered at the point $(\theta', \omega)$. The known fraction of oscillators $G'(\omega) d\omega$ in the interval $d\omega$ is given in terms of the distribution function $f$ by $G'(\omega) = (2\pi)^{-1} \int d\theta' f(\theta', \omega)$. The evolution of the distribution function is determined by the continuity equation,

$$\frac{\partial}{\partial t} f + \frac{\partial}{\partial \theta'}\left[\left(\omega + \mathrm{Im}(F e^{-i\theta'})\right) f\right] = 0,  \tag{2}$$

where the complex force $F$ is the sum of the coherent response of the ensemble of oscillators, and the externally imposed driving term.

$$F = K \int \frac{d\theta' d\omega}{2\pi} f e^{i\theta'} + M e^{i\Omega t}.  \tag{3}$$

Here $M = M_0 e^{i\alpha}$ is the complex drive amplitude.

We now seek solutions of the system (2) and (3) in terms of an equilibrium state and small perturbations away from that equilibrium state. Here by 'equilibrium' we mean that the time dependence of the force $F$ is an oscillation purely at the same frequency $\Omega$ as the drive term (i.e., the external force entrains the system of oscillators). The equilibrium will be time independent in a rotating frame that we obtain under the transformation $\omega \to \delta\omega = \omega - \Omega$, and $\theta' \to \theta + \Omega t$. This transformation leaves the form of Eqs. (2) and (3) unchanged, except that the drive term is now time independent. The PDF of $\delta\omega$ is $G(\delta\omega) = G'(\delta\omega + \Omega)$, i.e., the PDF in $\delta\omega$ is shifted by $\Omega$. In the rotating frame we write both the force $F$ and the distribution function $f$ as the sum of a steady part and a small time dependent perturbation, $F(t) = F_0 + F_1(t)$ and $f = f_0(\delta\omega, \theta) + f_1(\delta\omega, \theta, t)$. We note that the force $F$ is a complex variable. Without loss of generality, we can choose the phase $\alpha$ of the drive term $M$, so that the equilibrium force $F_0$ is real. The perturbed force $F_1$ will remain complex, and the phase of $M$ must be determined as part of the equilibrium.

**A. Equilibrium**

The lowest order, equilibrium, version of (2) takes the form,

$$\frac{\partial}{\partial \theta}\left[(\delta\omega - F_0 \sin\theta)f_0\right] = 0. \tag{4}$$

Solutions to Eq. (4) depend on whether an individual oscillator is trapped by the coherent force, $|\delta\omega| < F_0$, or whether its angle changes monotonically with time, $|\delta\omega| > F_0$. In the first case an oscillator's angle is attracted to a fixed angle $\theta_s(\delta\omega / F_0)$, where $\delta\omega = F_0 \sin\theta_s$, and the distribution function takes the form of a delta function at this angle. (Of the two possible angles $\theta_s$ satisfying the preceding relation, the one for which $\cos\theta_s > 0$ is attracting and is selected.) In the second case, the quantity in the square brackets in Eq. (4) is a constant, independent of angle. In both cases the form of the distribution function includes a multiplicative constant whose value is determined by the requirement that the angular average of the distribution function produces the known distribution of oscillators, $G(\delta\omega)$. Performing the required averages leads to,

$$f_0(\theta, \delta\omega) = G(\delta\omega)\begin{cases} \sqrt{\delta\omega^2 - F_0^2}/|\delta\omega - F_0 \sin\theta|, & |\delta\omega| > F_0 \\ 2\pi\delta(\theta - \theta_s(\delta\omega / F_0)), & |\delta\omega| \leq F_0 \end{cases}. \tag{5}$$

We insert the expression (5) for the equilibrium distribution function into Eq. (3) for the force and obtain,

$$F_0 - M_0 e^{i\alpha} = K\,R(\Omega, F_0), \tag{6}$$

where the complex equilibrium order parameter $R$ is given by,

$$R = \int \frac{d\theta d\delta\omega}{2\pi} f_0 e^{i\theta} = \int \frac{d\delta\omega}{F_0} G(\delta\omega) \begin{cases} i\delta\omega - i\sqrt{\delta\omega^2 - F_0^2}, & \delta\omega > F_0 \\ i\delta\omega + \sqrt{F_0^2 - \delta\omega^2}, & |\delta\omega| < F_0 \\ i\delta\omega + i\sqrt{\delta\omega^2 - F_0^2}, & \delta\omega < -F_0 \end{cases} \quad (7)$$

Equation (7) has the same form as is considered in the analysis of the coherent state of the undriven Kuramoto model. However, in that case it is frequently assumed that the distribution function is an even function of frequency. Then, the only term that survives the integral over frequency is the positive radical occurring for trapped oscillators $|\delta\omega| < F_0$, and the order parameter is real. In our case $G$ will, in general, not be even in $\delta\omega$, and all terms in (7) contribute [11]. Moreover, the order parameter will be a complex number reflecting the fact that it is not in phase with the force.

It is interesting to note that even though the distribution function (Eq. (5)) has very different forms depending on whether an individual oscillator's phase is trapped or not, the resulting contributions of oscillators to the order parameter after integration over angle (Eq. (7)) involve branches of the same function. In particular, defining $\sqrt{F_0^2 - \delta\omega^2}$ for complex $\delta\omega$ by the branch cuts shown in Fig. 1, we can express (7) as follows,

$$R = \int_{C_+} \frac{d\delta\omega}{F_0} G(\delta\omega)\left(i\delta\omega + \sqrt{F_0^2 - \delta\omega^2}\right), \quad (8)$$

where $C_+$ denotes a contour from $\delta\omega = -\infty$ to $\delta\omega = +\infty$ that is infinitesimally displaced upward from the real $\delta\omega$ axis as shown in Fig. 1.

The expression (8) for the order parameter is particularly useful if the distribution function of oscillator frequencies $G$, can be expressed as an analytic function of $\delta\omega$. An example that we will use here is that of a Lorentzian distribution,

$$G(\delta\omega) = \frac{1}{\pi\left[1 + (\delta\omega + \Omega)^2\right]} = \frac{1}{\pi(\delta\omega + \Omega + i)(\delta\omega + \Omega - i)}. \quad (9)$$

Without loss of generality, the Lorentzian distribution is chosen to have zero mean in the original reference frame, and a unit width. The choice of unit width serves to define the frequency scale in the problem. The integral in Eq. (8) can now be evaluated analytically by the method of residues. Closing the integration contour with a large semicircle in the upper half $\delta\omega$-plane and letting its radius approach infinity, we find that the value of the integral is determined by the residue at the pole $\delta\omega = -\Omega + i$. The order parameter $R$ is given by

$$R(\Omega, F_0) = \sqrt{1 + \left(\frac{1 + i\Omega}{F_0}\right)^2} - \frac{1 + i\Omega}{F_0}. \quad (10)$$

Solutions of the equilibrium Eq. (6) are most easily found by fixing the coupling strength $K$ and frequency $\Omega$, and plotting the drive amplitude, $M_0 = |M|$, as a function of the equilibrium force, $F_0$. Such plots are shown in Fig. 2 for several values of coupling and frequency. The continuous curves in Fig. 2 represent solutions of Eq. (6). The portions of the curves that are solid will be shown in the subsequent analysis to be stable equilibria, and the portions that are dashed are unstable. The symbols represent the results of numerical simulations of Eq. (1) for an ensemble of N=1000 oscillators with frequencies determined from the PDF $G(\delta\omega)$ using a random number generator. It should be noted that while the drive amplitude is a single valued function of force, if one regards the drive amplitude as specified, there can be multiple equilibrium force values. We find that for this distribution function there are either three equilibrium solutions (as in curves (b) and (c)) or one equilibrium solution (as in curve (a)) depending on the parameters $K$ and $\Omega$. In the limit of zero drive ($M_0 = 0$), solutions of Eqs. (6) and (10) yield the known result that there is a critical coupling value $K = 2$ such that the only solution for $K < 2$ is $R = 0$, while for $K > 2$, $\Omega = 0$ and $R = \sqrt{1 - 2/K}$. In the next section we determine the stability of the equilibria described by Eqs. (6) and (10).

**A. Stability**

Linear perturbation of Eqs. (2) and (3) about the equilibrium given by (5) results in equations for the first order distribution function,

$$\frac{\partial}{\partial t}f_1 + \frac{\partial}{\partial \theta}\left[(\delta\omega - F_0 \sin\theta)f_1\right] + \frac{\partial}{\partial \theta}\left[\frac{1}{2i}(F_1 e^{-i\theta} - F_1^* e^{+i\theta})f_0\right] = 0 \tag{11}$$

and the first order collective force,

$$F_1 = K\int \frac{d\theta' d\delta\omega}{2\pi} f_1 e^{i\theta} \quad . \tag{12}$$

To solve the system (11) and (12) we introduce Laplace transforms in time with transform variable $s$. Since (11) involves both $F_1$ and its complex conjugate we add to the system the conjugate of Eq. (12) and treat both $F_1$ and its conjugate as separate variables, each with its own Laplace transform. (In general, these transforms are not conjugates of each other, unless $s$ is real.) Further, since we are interested in the normal modes of (11) and (12) we neglect the initial condition on the perturbed distribution function and seek the characteristic equation for complex growth rate $s$ [12]. After Laplace transformation, Eqs. (11), (12) and the conjugate of (12) are replicated with each first order quantity replaced by its transform, and the time derivative in (11) replaced by $s$.

To solve the transformed equations we must separately consider the case of trapped phases, $|\delta\omega| \leq F_0$, and untrapped phases, $|\delta\omega| > F_0$. We therefore break the $\delta\omega$ integration in (12) into three intervals

$$F_1 = F_1^> + F_1^T + F_1^<, \tag{13}$$

where $F_1^>$ ($F_1^<$) is the contribution from untrapped oscillators with $\delta\omega > F_0$ ($\delta\omega < F_0$), and $F_1^T$ is the contribution of trapped oscillators $|\delta\omega| \leq F_0$. In the case of trapped phases, the unperturbed distribution function is in the form of a delta function of angle (see Eq. (5)) that falls under the derivative with respect to angle in the last term in Eq. (11). Using the identities,

$$d[g(\theta)\delta(\theta-\theta_s)]/d\theta = g(\theta_s)\delta'(\theta-\theta_s), \tag{14a}$$

and

$$h(\theta)\delta'(\theta-\theta_s) = h(\theta_s)\delta'(\theta-\theta_s) - h'(\theta_s)\delta(\theta-\theta_s), \tag{14b}$$

where prime indicates differentiation with respect to $\theta$, we find that the perturbed distribution function for trapped oscillators ($|\delta\omega| \leq F_0$) is given by,

$$\bar{f}_1 = i\pi \frac{G(\delta\omega)\,\delta'(\theta-\theta_s)}{s + \sqrt{F_0^2 - \delta\omega^2}} \left(\bar{F}_+ e^{-i\theta_s} - \bar{F}_- e^{+i\theta_s}\right), \tag{15}$$

where $\bar{F}_+$ and $\bar{F}_-$ are the Laplace transforms of $F_1$ and its complex conjugate, respectively. Equation (15) can be understood as follows. Oscillators with trapped phases are attracted to the angle $\theta_s$ giving rise to a distribution function that is a delta function of angle at $\theta_s$. Under perturbation, the oscillators are displaced from $\theta_s$, and the perturbed distribution takes the form of the negative of the displacement multiplying the derivative of a delta function. The denominator in (15) shows that the rate at which the displaced phases relax back to $\theta_s$ in the uncoupled case is given by $\sqrt{F_0^2 - \delta\omega^2}$.

Substituting Eq. (15) into Eq. (12) and its conjugate we arrive at expressions for the contribution of trapped oscillators to the perturbed force,

$$\bar{F}_+^T = \frac{1}{2}K \int_{|\delta\omega|\leq F_0} \frac{d\delta\omega\, G(\delta\omega)}{s + \sqrt{F_0^2 - \delta\omega^2}} \left(\bar{F}_+ - \bar{F}_- e^{2i\theta_s}\right), \tag{16a}$$

and

$$\bar{F}_-^T = \frac{1}{2}K \int_{|\delta\omega|\leq F_0} \frac{d\delta\omega\, G(\delta\omega)}{s + \sqrt{F_0^2 - \delta\omega^2}} \left(\bar{F}_- - \bar{F}_+ e^{-2i\theta_s}\right). \tag{16b}$$

Expressions (16a) and (16b) include only the contributions of the trapped oscillators to $\bar{F}_\pm$. A separate analysis must be performed to determine the contributions of the untrapped oscillators, which must be included in Eq. (13). This analysis is involved but is presented in the Appendix. As in the case of the equilibrium, the contributions of the trapped and untrapped oscillators form different branches of the same function (c. f. Eq. (A13) of the Appendix). Equations (16a) and (16b) along with (A13a, b) and (A14a, b) of the appendix combined in (13) then yield

$$\bar{F}_+ = K\left(\chi_+^+ \bar{F}_+ + \chi_-^+ \bar{F}_-\right), \tag{17a}$$

and

$$\bar{F}_- = K\left(\chi_+^- \bar{F}_+ + \chi_-^- \bar{F}_-\right). \tag{17b}$$

Here the "susceptibilities" $\chi_\pm^\pm$ are given in terms of integrals over the complex variable $\delta\omega$

$$\chi_+^+ = \frac{1}{2} \int_{C_+} \frac{d\delta\omega\, G(\delta\omega)}{s + \sqrt{F_0^2 - \delta\omega^2}}, \tag{18a}$$

$$\chi_-^- = \frac{1}{2} \int_{C_-} \frac{d\delta\omega\, G(\delta\omega)}{s + \sqrt{F_0^2 - \delta\omega^2}}, \tag{18b}$$

$$\chi_-^+ = -\frac{1}{2F_0^2} \int_{C_+} \frac{d\delta\omega\, G(\delta\omega)}{s + \sqrt{F_0^2 - \delta\omega^2}} \left(i\delta\omega + \sqrt{F_0^2 - \delta\omega^2}\right)^2, \tag{18c}$$

and

$$\chi_+^- = -\frac{1}{2F_0^2} \int_{C_-} \frac{d\delta\omega\, G(\delta\omega)}{s + \sqrt{F_0^2 - \delta\omega^2}} \left(-i\delta\omega + \sqrt{F_0^2 - \delta\omega^2}\right)^2. \tag{18d}$$

Here $C_+$ is the contour defined in Fig. 1, $C_-$ is the analogous contour running from $\delta\omega = -\infty$ to $\delta\omega = +\infty$ along a path displaced infinitesimally *downward* from the real $\delta\omega$-axis, and we have again used the branch cuts in Fig. 1 to define $\sqrt{F_0^2 - \delta\omega^2}$.

The characteristic equation for the complex growth rate s is obtained by solving the simultaneous equations (17a) and (17b),

$$\left(1 - K\chi_+^+\right)\left(1 - K\chi_-^-\right) - K^2 \chi_+^- \chi_-^+ = 0. \tag{19}$$

We note that in the special case of an even PDF $G(\delta\omega) = G(-\delta\omega)$ the susceptibilities have the following relations: $\chi_+^- = \chi_-^+$, $\chi_+^+ = \chi_-^-$, and (19) reduces to two uncoupled equations, $1 = K(\chi_+^+ + \chi_+^-)$, and $1 = K(\chi_+^+ - \chi_+^-)$ that agree with the characteristic equations of Ref. 10.

The "susceptibilities" $\chi_\pm^\pm$ given by Eqs. (18a-18d) can be evaluated by the method of residues for the Lorentzian distribution function given by (9). In the case of the contour $C_+$ ($C_-$) the integral is closed using a large semicircle in the upper (lower) half plane, and only the pole at $\delta\omega = -\Omega + i$ ($\delta\omega = -\Omega - i$) is enclosed by the contour. Thus, relatively simple expressions for the susceptibilities are obtained,

$$\chi_+^+ = \frac{1}{2\left(s + F_0\sqrt{1+\beta^2}\right)}, \tag{20a}$$

$$\chi_-^+ = -\frac{\left(\sqrt{1+\beta^2} - \beta\right)^2}{2\left(s + F_0\sqrt{1+\beta^2}\right)}, \tag{20b}$$

$$\chi_-^- = \frac{1}{2\left(s + F_0\sqrt{1+\beta^{*2}}\right)}, \tag{20c}$$

and

$$\chi_+^- = -\frac{\left(\sqrt{1+\beta^{*2}} - \beta^*\right)^2}{2\left(s + F_0\sqrt{1+\beta^{*2}}\right)}, \tag{20d}$$

where $\beta = (1 + i\Omega)/F_0$, and $\beta^*$ denotes the complex conjugate of $\beta$. Inserting expressions (20a)-(20d) for the "susceptibilities" into Eq (19) results in a quadratic equation for $s$,

$$\left(s + F_o\sqrt{1+\beta^2} - \frac{K}{2}\right)\left(s + F_o\sqrt{1+\beta^{*2}} - \frac{K}{2}\right) = \left[\frac{K\left(\sqrt{1+\beta^2} - \beta\right)\left(\sqrt{1+\beta^{*2}} - \beta^*\right)}{2}\right]^2. \tag{21}$$

If one multiples out the terms in Eq. (21), one sees that the coefficients in the quadratic equation for $s$ are all real. Consequently, there will be either two real solutions for $s$, or a pair of complex conjugate solutions.

We have solved (21) for the equilibria displayed in Fig. 2, and indicated the range of $F_0$ values where both solutions are stable (solid lines) and unstable (dashed lines). For curve (c) corresponding to $K=7$ and $\Omega = 2$ the behavior of the roots of (21) as $F_0$ is varied is as follows. For $F_0 > 6.2$ corresponding to values greater than the local minimum of the curve $M_0$ vs. $F_0$, there are two real, stable solutions of (21). For $F_0$ between 2.7 and 6.2 corresponding to the local maximum and minimum there are two real solutions of (21) with one stable and one unstable. For $F_0$ between 2.27 and 2.7 ($F_0 = 2.7$ corresponds to the local maximum of $M_0$) the two real solutions are both unstable. In fact, the existence of an extremum in the $M_0$ vs. $F_0$ curve implies a perturbation solution with $s=0$. As $F_0$ is decreased through $F_0 = 2.27$ the two real, unstable solutions merge and become complex conjugate, unstable solutions.

# III PARAMETER SPACE FOR LOCKED STATES

Figure 2 shows that depending on parameters there may be one or three equilibrium locked states for a given coupling strength $K$, amplitude $M_0$ and frequency $\Omega$ of the drive signal. In this section we describe the range of parameters for which different numbers of stable and unstable equilibria exist, and the nature of the transitions between these states as parameters are varied.

For $K>2$ (i.e., above the threshold for the appearance of the coherent state in the undriven system) and for frequency below some critical value, the drive amplitude versus force curve is nonmonotonic as shown by the $K=5$ and $K=7$ curves of Fig. 2. For $K<2$ or for $K>2$ and for frequencies above the critical value the drive amplitude versus force curve is monotonic as illustrated by the $K=1$ curve of Fig. 2. This means that for a given $K$ the $M_0$ vs. $\Omega$ plane will be divided into regions where there are either one or three equilibria. Further, analysis of the stability of these equilibria using Eq. (21) shows that the plane can be divided into three regions where there are A) one stable, B) one unstable, or C) one stable and two unstable equilibria. This is illustrated in Fig. 3 for the case $K=5$, where the three regions are separated by solid lines. Region A corresponds to values of drive amplitude and frequency giving one stable equilibrium, region B corresponds to values giving one unstable equilibrium, and region C corresponds to values giving one stable and two unstable equilibria. In terms of Fig. 2, region C corresponds to values of drive amplitude between the local maximum and minimum of the $M_0$ vs. $F_0$ curve. The point labeled T ($\Omega = 3.54$, $M_0 = 3.41$) corresponds a value of frequency above which the $M_0$ vs. $F_0$ curve increases monotonically, and only a single equilibrium is possible. The boundary separating regions A and B is defined by the stability of that single equilibrium. Regions A and C are further subdivided by the dotted line which gives the parameters for which the two solutions for $s$ merge. Above this curve all solutions for $s$ are real, while below it there are complex conjugate pairs of solutions. In the case of region A the merged solutions are stable, while in region C the merged solutions are unstable (corresponding to the equilibrium with the smallest value of $F_0$ in curve (b) of Fig. 2). This division of parameter space depicted in Fig. 3 can be thought of as a generalization of the Adler relation [13] to a system of phase oscillators. In particular, for a given drive amplitude, there is a maximum frequency (determined by the boundary of region B) above which the system will not lock to the drive signal, and the larger the drive amplitude, the greater the range of frequencies for which the system of oscillators can be locked.

Our linear analysis has shown that there are generally two independent normal mode perturbations that satisfy the characteristic equation (21) for each equilibrium solution. This suggests that the dynamics around the equilibria will be analogous to those of a two-dimensional autonomous system. This cannot be true in general as the original system is $N$ dimensional, and we are considering large values of $N$. The full spectrum of perturbations of the system includes the discrete normal modes of Eq. (21) as well as a continuous spectrum [12]. As the continuous spectrum is stable or marginally stable, we expect the two-dimensional analogy to hold as long as one is not too near a boundary in Fig. 3 where the normal modes are marginally stable. Here we describe transitions between different regions based on the two dimensional picture.

A schematic diagram illustrating the behavior of the system for different parameter values is shown in Fig. 4, where phase portraits are plotted for each of the five regions of Fig. 4. One can think of the phase portraits as showing the evolution of the system in the complex plane of the force $F$ appearing in Eq. (3). In region A there is a single stable equilibrium corresponding to a locked state. In region $A_1$ perturbations of this equilibrium have two real solutions for $s$, whereas in region $A_2$ the solutions for $s$ are complex conjugates. In region B there is a single unstable equilibrium and an attracting limit cycle. Motion on the limit cycle represents the tendency for the system to follow the undriven coherent state in cases in which the driving amplitude is too weak or the driving frequency to large for phase locking to occur. The transition from region A to region B is *analogous* to a super-critical Hopf bifurcation where the stable equilibrium becomes unstable and a periodic orbit appears [14]. In region C there are three equilibrium solutions, labeled 1, 2, and 3. In region $C_1$ equilibrium 1 is stable with two negative real values of $s$, equilibrium 2 is unstable with one positive and one negative real value of $s$, and equilibrium 3 is unstable with two real values of $s$. The region $C_2$ is similar except that equilibrium 3 is unstable with complex conjugate values of $s$. In the transition between region $A_1$ and $C_1$ the stable equilibrium of region $A_1$ becomes equilibrium 1 in region $C_1$, and equilibria 2 and 3 are created. This is analogous to a saddle-node bifurcation. When the boundary between region $C_1$ and $C_2$ is crossed equilibrium 3 changes from having two real positive values of $s$, to two complex conjugate unstable values of $s$. When the boundary between regions C and B is crossed, equilibrium solutions 1 and 2 merge leaving the limit cycle. The transition from B back to $C_2$ is analogous to a saddle- node bifurcation on a periodic orbit.

We note that the parameter space of Figs. 3 and 4 is organized by the critical point T. At this point the two solutions of Eq. (21) for $s$ merge at $s=0$. This can be seen by considering the curves of Fig. 2. The existence of an extremum in the $M_0$ vs. $F_0$ curve implies the existence of a zero frequency ($s=0$) perturbation of the equilibrium at the extremum. At this point a small change in $F$ is accompanied by no change in $M_0$. As the $M_0$ vs. $F_0$ curve changes from being nonmonotonic to monotonic the two extrema merge at which point both solutions of (21) have $s=0$. A plot of the frequency and drive coordinates of this point as a function of coupling strength $K$ is shown in Fig. 5.

The asymptotic time dependence of
$$r = N^{-1}\sum e^{i\theta_j} ,$$
evaluated in the frame rotating at frequency $\Omega$ is described as follows. In regions A and C it is steady in time, while in region B it varies periodically in time. This is illustrated in Fig. 6, where we show power spectra of $r(t)$ for four representative points in the plane of Fig. 4. In displaying the power spectra we have shifted back to the original reference frame in which a state locked to the drive appears as a single peak at frequency $\Omega$ and the coherent state in the absence of drive appears as a single peak at zero frequency. The interesting spectra correspond to the two parameter values falling in region B. Figure

6B1 shows the spectrum for a point in Region B close to the boundary with region A. At this boundary the phase locked state is marginally stable and the solutions of Eq. (21) are complex conjugates corresponding to modes with real frequency. As a result, the power spectrum at this point shows a peak at the driving frequency and satellite peaks at a set of frequencies uniformly spaced from the driving frequency by approximately the real frequency of the marginally stable modes. In contrast, Fig. 6B2 shows the spectrum at a point in region B close to the boundary with region C. Here the spectrum also has a peak at the drive frequency and a set of uniformly spaced satellites surrounding the drive frequency. However, in this case the spacing of the satellites is small. This can be understood based on the phase portraits of Fig. 4. At the boundary between regions B and C the two equilibria labeled 1 and 2 in region $C_2$ of Fig. 4 merge leaving a stable limit cycle. Close to the boundary in region B the vestige of the two equilibria is felt on the limit cycle. In particular, as a phase point travels around the limit cycle, it spends a long time close to the location of the point where equilibria 1 and 2 have merged. Thus, at the boundary the period of the limit cycle becomes infinite, and the frequency spacing of the satellites goes to zero. In our simulations of this case we find that there is a significant finite $N$ dependence for parameters near this transition. In particular, if the number of oscillators is increased from 1000 to 5000 there is a decrease in the spacing of the peaks for the case of Fig. 6B2. This can be explained as follows. The effect of a finite number of oscillators can be mimicked in the infinite $N$ system by adding a noise source to the force $F$. The effect of this noise will be to decrease the time the phase point spends near the vestige of the two equilibrium points 1 and 2. Thus, the frequency spacing for a simulation with a larger number of oscillators and lower noise level will be smaller than that of a simulation with a lower number of oscillators. We leave a quantification of this effect for future study.

## IV CONCLUSIONS

We have analyzed the driven Kuramoto model, which can be considered as describing a wide class of phenomena in which systems of coupled oscillators can be entrained by an externally imposed driving force. Our analytic approach has been to search for equilibrium solutions in which the oscillators respond coherently to the driving force, and to analyze the linear stability of these equilibrium solutions. The analysis has been complemented and supported by numerical simulations.

The model has essentially four independent parameters: the number of oscillators $N$, the strength of the oscillator-to-oscillator (all-to-all) coupling $K$, and the amplitude $M_0$ and frequency $\Omega$ of the drive frequency. (The system of oscillators is assumed to have a distribution of natural frequencies with zero mean and unit width, and this serves to define a natural frequency scale in the problem). Our analysis focused on the $N \to \infty$ limit. We obtained a transcendental equation for the equilibrium state in terms of the parameters and a contour integral over individual oscillator frequencies. For the special case of a Lorentzian distribution of oscillator frequencies we evaluated the contour integral by the method of residues and obtained an algebraic equation relating the equilibrium parameters. For a given set of parameters we found one or three equilibrium solutions.

The stability analysis revealed that for our chosen distribution function an equilibrium would have two normal mode solutions for growth rate *s*, which were either both real, or were a complex conjugate pair. Knowledge of the stability of the equilibria allowed us to determine the regions of parameter space where locked states would be found. This knowledge also led us to describe the nature of the bifurcations of the equilibria as parameters were varied. Specifically we found that the transition between a locked and an unlocked state could be analogous to either a super-critical Hopf bifurcation or a saddle-node bifurcation.

An open question requiring further study is the detailed nature of the transitions taking into account the high dimensional nature of the system. Our analysis based on normal modes suggested that the system behavior was similar to that of a two-dimensional autonomous system. However, there are also modes of the system that have a continuous spectrum. These could give rise to higher dimensional behavior close to the stability boundaries of the locked states. So far however, our numerical simulations suggest that the two-dimensional picture is quite accurate.

We note that in a separate paper we will show that there is a special set of distribution functions [15] that form an infinite dimensional submanifold $\mathcal{M}$ in the space $\mathcal{B}$ of all possible distribution functions such that for any initial condition $f(\theta, \omega, t = 0)$ on $\mathcal{M} \subset \mathcal{B}$

> *(i)* the solution, $f(\theta, \omega, t)$, of Eq. (2) and (3) remains on $\mathcal{M}$ [i.e. $\mathcal{M}$ is invariant]; and
> *(ii)* the evolution of the complex force function *F(t)* is two dimensional [i.e., there is a system of two real first order nonlinear ordinary differential equations governing the evolution of *F(t)*].

In particular, this establishes that the nonlinear picture we have given in Fig. (3) applies on this submanifold. Furthermore, our numerical experiments suggest that this submanifold is attracting (our simulations initially placed oscillators at random in the $(\theta, \omega)$ plane). We also wish to emphasize that, although the dynamics for $f(\theta, \omega, t)$ on $\mathcal{M}$ is infinite dimensional, the dynamics for the macroscopic quantity *F(t)* given by integration over $(\theta, \omega)$ in Eq. (3) is exactly two-dimensional. Thus, while the linearized problem for *f* on $\mathcal{M}$ has a continuous spectrum, this does not lead to infinite dimensional dynamics of *F(t)*.

## V ACKNOWLEDGEMENTS

The authors acknowledge the helpful comments of Steven Strogatz. This work was supported by grants from the National Science Foundation and the Office of Naval Research.

## APPENDIX: LINEAR RESPONSE OF UNTRAPPED OSCILLATORS

In this appendix we outline the solution of the linearized continuity equation (11) for the case of oscillators whose phases are not trapped by the equilibrium force $F_0$, and we obtain explicit expressions for $\bar{F}_\pm^>$ and $\bar{F}_\pm^<$. For clarity of presentation we consider only the case $\delta\omega > F_0$, the result can then be generalized to the case $\delta\omega < -F_0$. For untrapped oscillators the phase angle in equilibrium increases at a rate given by, $\dot{\theta}_0 = \delta\omega - F_0 \sin\theta$. To solve (11) we introduce the time of flight required to travel from $\theta = 0$ to some specified angle $\theta$,

$$\tau(\theta) = \int_0^\theta \frac{d\theta}{\dot{\theta}_0}. \tag{A1}$$

In equilibrium, the motion of these oscillators is periodic with a period $T$ given by

$$T(\delta\omega) = \int_0^{2\pi} \frac{d\theta}{\dot{\theta}_0} = \frac{2\pi}{\sqrt{\delta\omega^2 - F_0^2}}. \tag{A2}$$

The integral in Eq. (A1) can be evaluated, and the transformation inverted giving the angle in terms of the travel time. Specifically, $e^{i\theta} = \varsigma(\tau)$, where

$$\frac{\varsigma - \varsigma_-}{\varsigma - \varsigma_+} = \frac{1 - \varsigma_-}{1 - \varsigma_+} e^{2\pi i\tau/T} \equiv \rho e^{2\pi i\tau/T}, \tag{A3}$$

and $\varsigma_\pm = i\left(\delta\omega \pm \sqrt{\delta\omega^2 - F_0^2}\right)/F_0$. We note that using the expressions for $\varsigma_\pm$ we conclude $|\rho| < 1$.

Under the transformation (A1)-(A3) the linearized and Laplace transformed continuity equation (11) becomes

$$s\left(\dot{\theta}_0 \bar{f}_1\right) + \frac{d}{d\tau}\left(\dot{\theta}_0 \bar{f}_1\right) + \frac{d}{d\tau}\left[\frac{\Gamma(\delta\omega)}{2i\dot{\theta}_0}\left(\bar{F}_+ e^{-i\theta} - \bar{F}_- e^{+i\theta}\right)\right] = 0, \tag{A4}$$

where $\Gamma(\delta\omega) = \sqrt{\delta\omega^2 - F_0^2} G(\delta\omega)$, and we have used Eq. (5) to express the equilibrium distribution function. When viewed as functions of $\tau$, all quantities in (A4) are periodic with period $T$. We then write each term as a Fourier series,

$$\left(\dot{\theta}_0 \bar{f}_1\right) = \sum_m f_m e^{2\pi im\tau/T} \tag{A5}$$

$$\frac{\left(\bar{F}_+ e^{-i\theta} - \bar{F}_- e^{+i\theta}\right)}{\dot{\theta}_0} = \sum_m \left(\bar{F}_+ g^*_{-m} - \bar{F}_- g_m\right) e^{2\pi i m \tau / T}, \tag{A6}$$

where

$$g_m = \int_0^T \frac{d\tau}{T} \frac{e^{i\theta}}{\dot{\theta}_0} e^{-2\pi i m \tau / T}. \tag{A7}$$

Substituting (A5)-(A7) into (A4) and Fourier transforming we find

$$f_m = -\frac{m\bar{\omega}\Gamma}{2(s+im\bar{\omega})}\left(\bar{F}_+ g^*_{-m} - \bar{F}_- g_m\right), \tag{A8}$$

where $\bar{\omega} = 2\pi/T$ is the average rotational rate in the presence of the coherent force.

We now determine the coefficients $g_m$ defined in Eq. (A7). We convert the time integral to an integral around the unit circle of the complex variable $z = e^{2\pi i \tau/T}$. In doing this we use the transformation (A3) to express $e^{i\theta} = \varsigma$ as a function of $z$. The result is

$$g_m = \oint \frac{dz}{2\pi i z} z^{-m} \frac{\varsigma(z)}{\delta\omega - F_0(\varsigma - \varsigma^{-1})/(2i)}, \tag{A9}$$

which can be evaluated by the method of residues. The only nonzero terms are $g_{\pm 1}$ and $g_0$, the latter of which does not contribute to the perturbed distribution function according to (A8). The expressions for $g_{\pm 1}$ are,

$$g_1 = -\frac{2i}{F_0} \frac{\varsigma_+^2 \rho}{\left(\varsigma_+ - \varsigma_-\right)^2}, \tag{A10a}$$

and

$$g_{-1} = -\frac{2i}{F_0} \frac{\varsigma_-^2 \rho^{-1}}{\left(\varsigma_+ - \varsigma_-\right)^2}. \tag{A10b}$$

Consequently, our expression for the perturbed distribution function (A5) only contains two terms rather than an infinity of terms. (Furthermore, any initial condition of the perturbed distribution function having $|m| > 1$ will evolve unaffected by the coherent force. These perturbations will oscillate with a frequency $m\bar{\omega}(\delta\omega)$ and form a continuous spectrum.)

Equation (A8) gives the perturbed distribution function in terms of the perturbed force; to obtain the characteristic equation we need to use (12) and its conjugate to evaluate the perturbed force in terms of the perturbed distribution function. Specifically, the contribution to $\bar{F}_\pm$ from untrapped oscillators with frequencies $\delta\omega > F_0$ is

$$\bar{F}_\pm^> = K \int_0^{2\pi}\int_{F_0}^\infty \frac{d\theta\, d\delta\omega}{2\pi} \sum_m \frac{f_m}{\dot{\theta}_o} e^{2\pi i m\tau/T} e^{\pm i\theta} = K \sum_m \int_{F_0}^\infty \frac{T d\delta\omega}{2\pi} f_m d_m^\pm, \qquad (A11)$$

where

$$d_m^\pm = \int_0^T \frac{d\tau}{T} e^{2\pi i m\tau/T} e^{\pm i\theta} = \oint \frac{dz}{2\pi i z} z^m \varsigma^{\pm 1}. \qquad (A12)$$

The integrals in (A12) may be evaluated by the method of residues. The relevant results are: $d_{-1}^+ = (\varsigma_- - \varsigma_+)\rho$, $d_1^- = (\varsigma_- - \varsigma_+)/(\rho\varsigma_+^2)$, and $d_1^+ = d_{-1}^- = 0$.

Putting the pieces together we have,

$$\bar{F}_+^> = K\frac{1}{2} \int_{F_0}^\infty \frac{d\delta\omega\, G(\delta\omega)}{s - i\sqrt{\delta\omega^2 - F_0^2}} \left(\bar{F}_+ - \varsigma_-^2 \bar{F}_-\right), \qquad (A13a)$$

and

$$\bar{F}_-^> = K\frac{1}{2} \int_{F_0}^\infty \frac{d\delta\omega\, G(\delta\omega)}{s + i\sqrt{\delta\omega^2 - F_0^2}} \left(\bar{F}_- - \varsigma_-^2 \bar{F}_+\right). \qquad (A13b)$$

Repeating the calculation for $\delta\omega < -F_0$, we obtain

$$\bar{F}_+^< = K\frac{1}{2} \int_{-\infty}^{-F_0} \frac{d\delta\omega\, G(\delta\omega)}{s + i\sqrt{\delta\omega^2 - F_0^2}} \left(\bar{F}_+ - \varsigma_+^2 \bar{F}_-\right), \qquad (A14a)$$

and

$$\bar{F}_-^< = K\frac{1}{2} \int_{-\infty}^{-F_0} \frac{d\delta\omega\, G(\delta\omega)}{s - i\sqrt{\delta\omega^2 - F_0^2}} \left(\bar{F}_- - \varsigma_+^2 \bar{F}_+\right). \qquad (A14b)$$

Adding the contributions in Eqs. (13) results in the expressions (18).

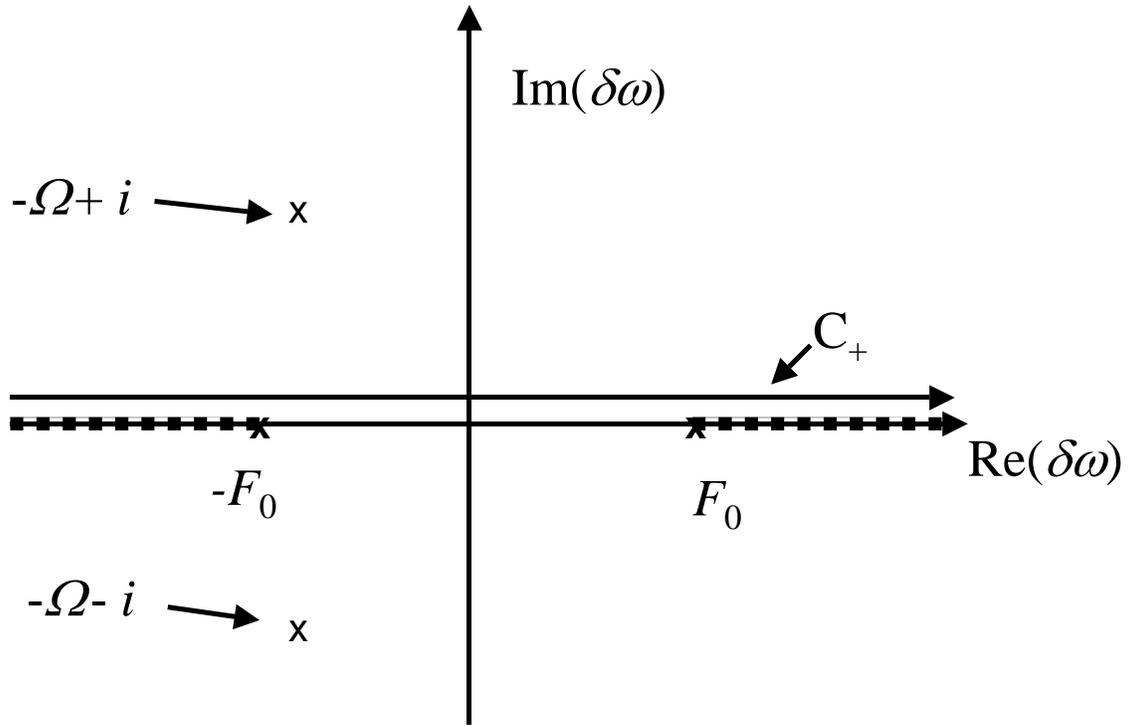

**Figure 1:** Integration path for Eq. (8), with branch cuts of $\sqrt{F_0^2 - \delta\omega^2}$ (shown as dashed lines) and poles of $G(\delta\omega)$ at $\delta\omega = -\Omega \pm i$.

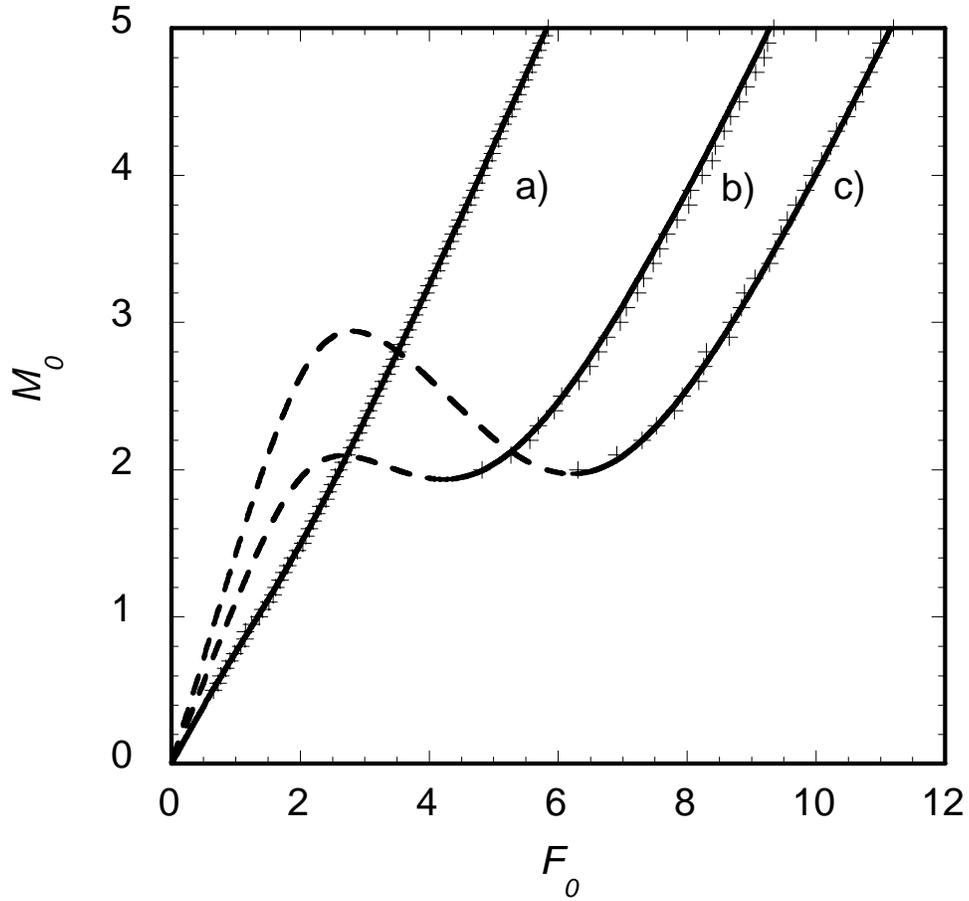

**Figure 2:** Driving strength $M_0$ vs. force $F_0$ for a) $K = 1$, $\Omega = 1$, b) $K = 5$, $\Omega = 2$, and c) $K = 7$, $\Omega = 2$. The continuous curves are solutions of Eq. (6) with stable equilibria (according to Eq. (21)) shown as solid lines and unstable equilibria shown as dashed lines. The crosses are the results of simulations of Eq. (1) with 1000 oscillators for the cases in which the force approached a steady value.

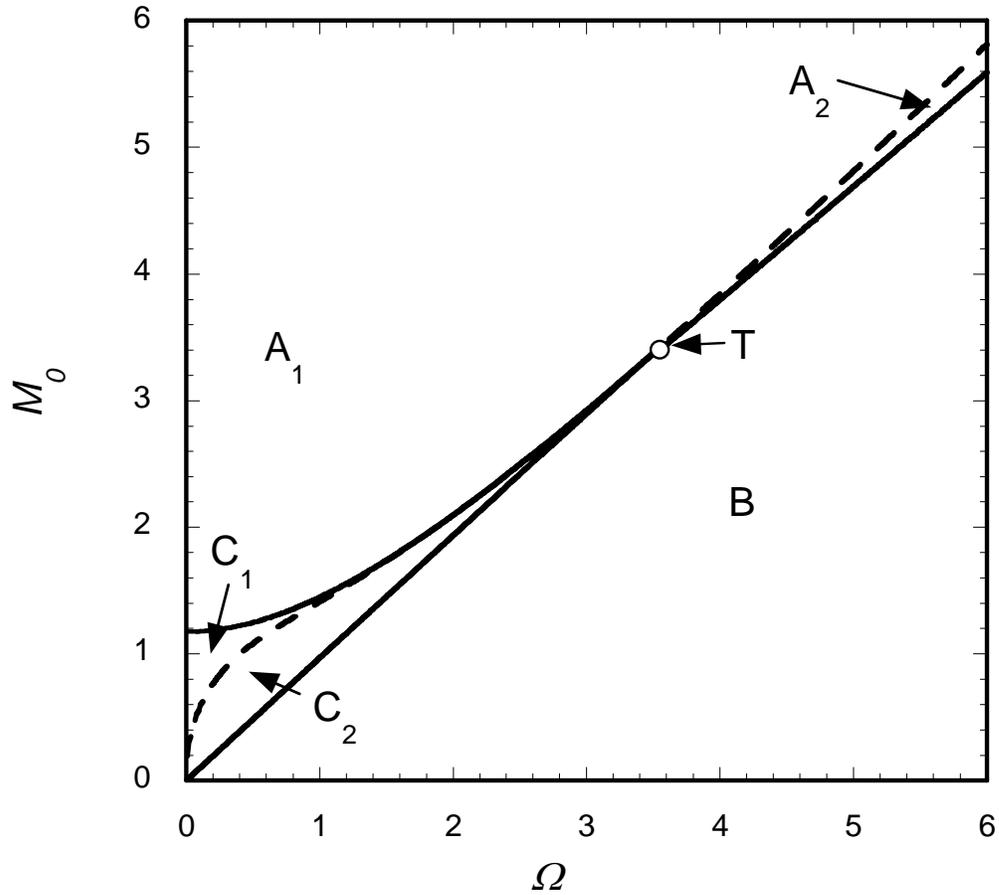

**Figure 3:** Parameter space of driving strength $M_0$ and frequency $\Omega$ for $K=5$ showing where different numbers of stable and unstable equilibria occur. In region A (B) there is one stable (unstable) equilibrium, and in region C there are one stable and two unstable equilibria. The dashed line corresponds to the set of points where the two solutions of Eq. (21) merge. At the point T the solid and dashed lines all intersect.

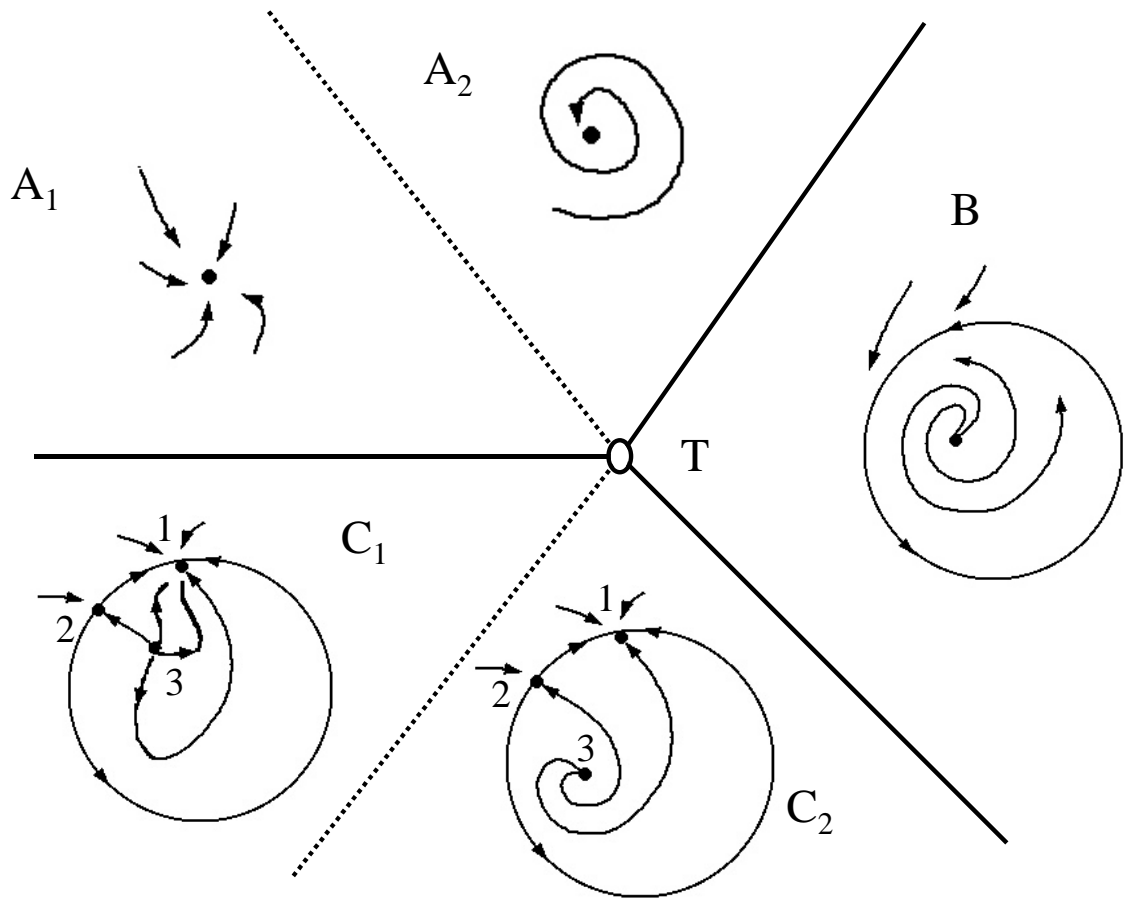

**Figure 4:** Schematic phase portraits of the equilibria in various parts of parameter space surrounding the point T.

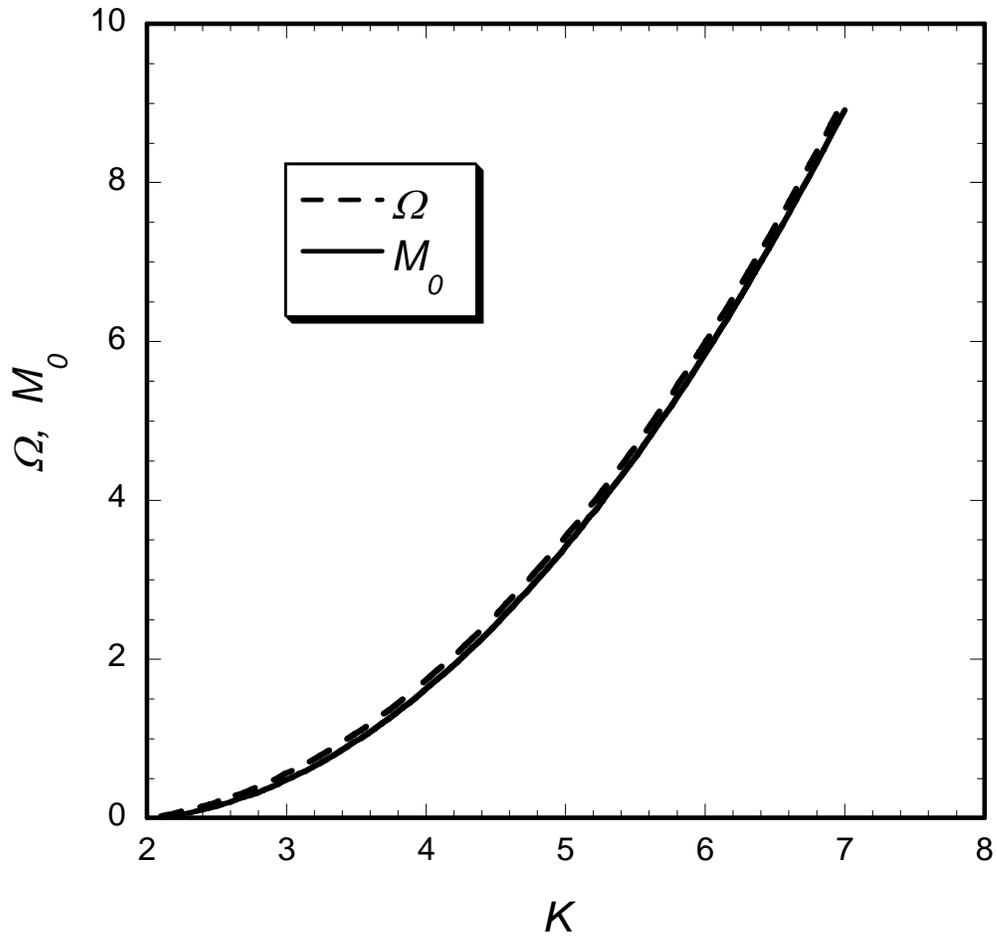

**Figure 5:** Coordinates of the critical point T as a function of the coupling strength *K*.

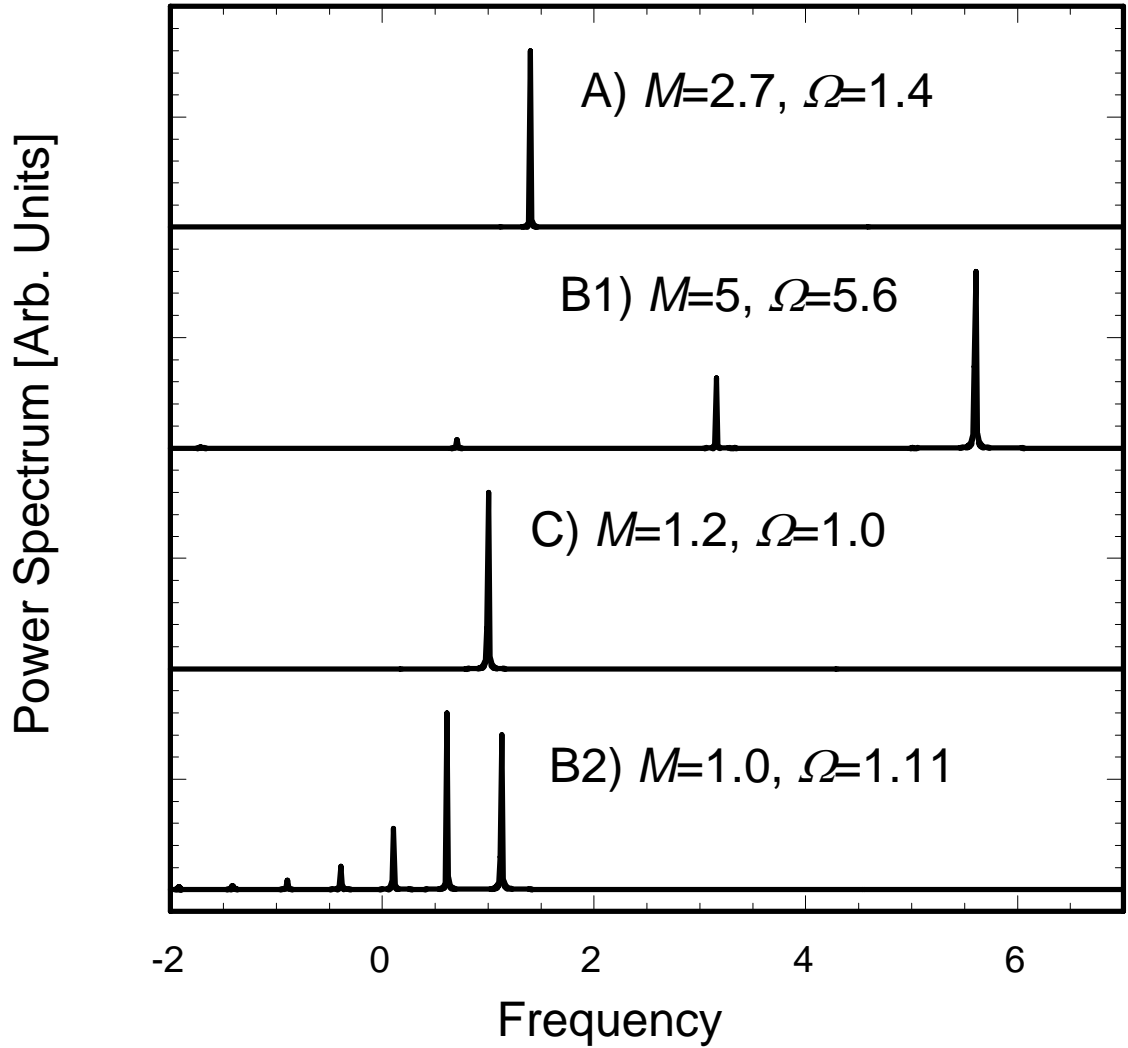

**Figure 6:** Power Spectra for *r(t)* for four representative points in the *M* vs. $\Omega$ plane for *K*=5. The four curves have been offset to separate them. The letter label of each curve indicates the corresponding region of Figs 3 and 4. Curve B1 is close to the boundary of region A, and curve B2 is close to the boundary of region C.